# Enhanced four-wave-mixing and 3rd order optical nonlinearity in SiN nanowires integrated with graphene oxide films

*Yang Qu, Jiayang Wu, Yunyi Yang, Yuning Zhang, Yao Liang, Houssein El Dirani, Romain Crochemore, Pierre Demongodinc, Corrado Sciancalepore, Christian Grillet, Christelle Monat, Baohua Jia,\* and David J. Moss\**

Y. Qu, Dr. J. Wu, Y. Zhang, Prof. D. J. Moss
Optical Sciences Centre,
Swinburne University of Technology,
Hawthorn, VIC 3122, Australia

Dr. Y. Yang, Yao Liang, Prof. B. Jia
Centre for Translational Atomaterials
Swinburne University of Technology
Hawthorn, VIC 3122, Australia

Dr. C. Sciancalepore, H.El Dirani, and R. Crochemore
University Grenoble Alpes,
CEA-LETI, Minatec, Optics ans Photonics Divesion,
17 rue des Martyrs, 38054 Grenoble, France
Current Address: Dr. C. Sciancalepore, SOITEC; H.El Dirani ST Microelectronics.

Dr. C. Grillet,
Institut des nanotechnologies de Lyon and
UMR CNRS 5270, Ecole Centrale Lyon,
F-69130 Ecully, France

Prof. C. Monat, P. Demongodin
Institut des nanotechnologies de Lyon, Ecole Centrale Lyon,
F-69130 Ecully, France

\*E-mail: bjia@swin.edu.au, dmoss@swin.edu.au

Keywords: 2D materials, graphene oxide, Kerr nonlinearity, four-wave mixing, integrated optics.

**Abstract**

Layered 2D graphene oxide (GO) films are integrated with silicon nitride (SiN) waveguides to experimentally demonstrate an enhanced Kerr nonlinearity via four-wave mixing (FWM). Owing to the strong light–matter interaction between the SiN waveguides and the highly nonlinear GO films, the FWM performance of the hybrid waveguides is significantly improved. SiN waveguides with both uniformly coated and patterned GO films are fabricated based on a transfer-free, layer-by-layer GO coating method together with standard photolithography and lift-off processes, yielding precise control of the film thickness, placement and coating length. Detailed FWM measurements are carried out for the fabricated devices with different numbers of GO layers and at different pump powers. By optimizing the trade-off between the nonlinearity and loss, we obtain a significant improvement in the FWM conversion efficiency of ≈7.3 dB for a uniformly coated device with 1 layer of GO and ≈9.1 dB for a patterned device with 5 layers of GO. We also obtain a significant increase in FWM bandwidth for the patterned devices. A detailed analysis of the influence of pattern length and position on the FWM performance is performed. Based on the FWM measurements, the dependence of GO's third-order nonlinearity on layer number and pump power is also extracted, revealing interesting physical insights about the 2D layered GO films. Finally, we obtain an enhancement in the effective nonlinear parameter of the hybrid waveguides by over a factor of 100. These results verify the enhanced nonlinear optical performance of SiN waveguides achievable by incorporating 2D layered GO films.

# 1. Introduction

The third-order optical nonlinearity ($\chi^{(3)}$), describing four-wave mixing (FWM), self-phase modulation (SPM), third harmonic generation (THG) and other effects [1- 12], has formed the basis for all-optical signal generation and processing that have achieved far superior performance in speed and operation bandwidth than electronic approaches [13-15]. As a fundamental $\chi^{(3)}$ process, FWM has found a wide range of applications in wavelength conversion [6, 7] , optical frequency comb generation [18, 19], optical sampling [20, 21], quantum entanglement [22, 23] and many others [24, 25]. Implementing nonlinear photonic devices in integrated form offers the greatest dividend in terms of compact footprint, high stability, high scalability and mass-producibility [1, 2, 26]. Although silicon has been a leading platform for integrated photonic devices for many reasons [1], including the fact that it leverages the well-developed complementary metal-oxide-semiconductor (CMOS) fabrication technologies [27], its strong two-photon absorption (TPA) at near-infrared telecommunications wavelengths poses a fundamental limitation for devices operating in this wavelength region. Other CMOS compatible platforms such as silicon nitride (SiN) and doped silica [2, 28] have a much lower TPA, although they still suffer from intrinsic limitation arising from a much lower Kerr nonlinearity.

The increasing demand for high performing nonlinear integrated photonic devices has motivated the search for highly nonlinear materials [20, 29]. The superior Kerr nonlinearity of 2D layered materials such as graphene, graphene oxide (GO), black phosphorus and transition metal dichalcogenides (TMDCs) has been widely recognized and has enabled diverse nonlinear photonic devices with high performance and new functionalities [29-40]. In particular, tunable FWM in graphene-covered SiN waveguides has been demonstrated by electronically tuning the Fermi energy of graphene [31].

Owing to its ease of preparation and the tunability of its material properties, GO has received increasing interest as a promising member of the 2D material family [41-46]. Previously, we reported GO films with a giant Kerr nonlinearity ($n_2$) of about 5 orders of magnitude higher than SiN [42], and demonstrated enhanced FWM in doped silica waveguides and microring resonators (MRRs) integrated with GO films [32, 47]. Unlike graphene, which has a metallic behavior with zero bandgap, GO is a dielectric with a distinct bandgap of 2.1−2.4 eV [41, 48]. This results in material absorption that is over 2 orders of magnitude lower than graphene [32] as well as negligible TPA in the telecommunications band [48, 49], both of which are highly desired for many nonlinear applications such as FWM. Moreover, by using a large-area, transfer-free, layer-by-layer GO coating method along with standard lithography and lift-off processes, we achieved GO film coating on integrated photonic devices with highly precise control of film thickness, placement and coating length [50]. This overcomes a critical fabrication bottleneck in terms of layer transfer for 2D materials [51] and marks an important step towards the eventual manufacturing of integrated photonic devices incorporated with 2D layered GO films.

In this paper, we report the integration of 2D layered GO films onto SiN waveguides − a CMOS-compatible platform that has been widely used for integrated nonlinear optics [2]. By using our GO fabrication techniques, both uniformly coated and patterned GO films are integrated on SiN waveguides with precise control of the film thickness, placement and coating length. Benefiting from the strong light–matter interaction between the SiN waveguides and the GO films with an ultrahigh Kerr nonlinearity and a relatively low loss, significantly improved FWM performance of the hybrid waveguides is achieved. We perform FWM measurements for different numbers of GO layers and at different pump powers, achieving a FWM conversion efficiency (CE) enhancement of ≈7.3 dB for a uniformly coated device with 1 layer of GO and ≈9.1 dB for a patterned device with 5 layers of GO. Both an improved FWM CE and bandwidth are achieved for the patterned devices compared to the uniformly coated devices. The influence of pattern length and position on FWM performance is also analysed. By fitting the experimental results with theory, the dependence of the $n_2$ of the GO film on layer number and pump power is extracted, showing interesting physical insights about the evolution of the layered GO films from 2D monolayers towards quasi bulk-like behavior. Finally, we obtain an improvement in the effective nonlinear parameter ($\gamma$) of the hybrid waveguides by over a factor of 100. These results reveal the strong potential of integrating 2D layered GO films on SiN devices to improve the nonlinear optical performance.

## 2. Device fabrication and characterization

### 2.1 Device fabrication

**Figure 1a** shows the SiN waveguide integrated with a GO film, along with a schematic showing atomic structure of GO with different oxygen functional groups (OFGs) such as hydroxyl, epoxide and carboxylic groups. The fabrication process flow for the device in **Figure 1a** is provided in **Figure 1b**.

SiN waveguides with a cross section of 1.6 μm × 0.66 μm were fabricated via annealing-free and crack-free processes that are compatible with CMOS fabrication [52, 53]. First, a SiN layer was deposited via low-pressure chemical vapor deposition (LPCVD) in two steps, with a 370-nm-thick layer for each, so as to control strain and to prevent cracks. In order to produce high-quality films, a tailored ultra-low deposition rate (< 2 nm/ min) was used. Waveguides were then formed via a combination of deep ultraviolet lithography and fluorine-based dry etching that yielded exceptionally low surface roughness. Next, a 3-μm thick silica upper cladding layer was deposited via high-density plasma-enhanced chemical vapor deposition (HDP-PECVD) to avoid void formation. To enable the interaction between the GO films and the evanescent field leaking from the SiN waveguides, the silica upper cladding was removed using a perfectly selective chemical-mechanical planarization (CMP) that left the top surface of the SiN waveguides exposed in air, with no SiN consumption and no remaining topography.

Layered GO films were coated on the top surface of the chip by a solution-based method that yielded layer-by-layer film deposition, as reported previously [32, 48, 50]. Four steps for the in-situ assembly of monolayer GO films were repeated to construct multilayer films. Our GO coating approach, unlike the sophisticated transfer processes employed for coating other 2D materials such as graphene and TMDCs [36, 54, 55], enables transfer-free and high-uniformity GO film coating over large areas (e.g., 4-inch wafers [48]), with highly scalable fabrication processes and precise control of the number of GO layers (i.e., GO film thickness).

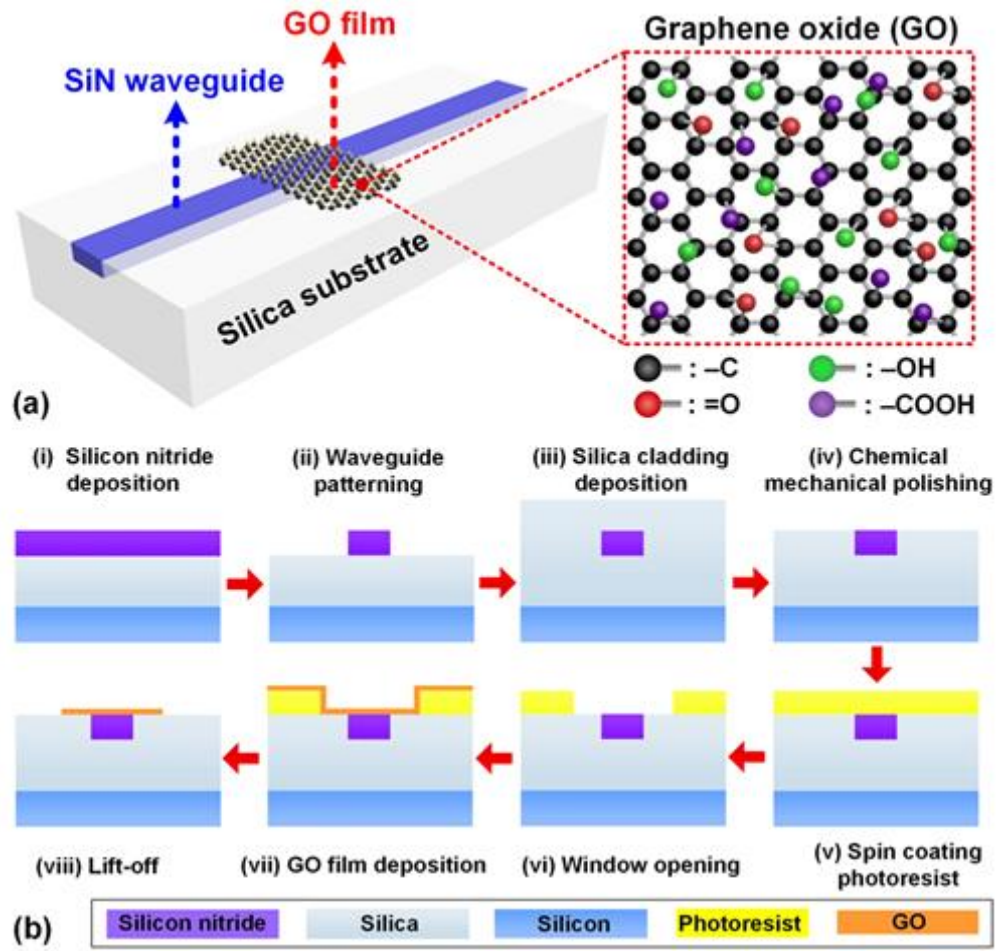

**Figure 1.** (a) Schematic illustration of GO-coated SiN waveguide. Inset shows the schematic atomic structure of GO. (b) Schematic illustration showing the fabrication process flow for the device in (a).

In addition to the uniformly coated devices, we selectively patterned GO films on SiN waveguides using standard lithography and lift-off processes. The chip was first spin-coated with photoresist and then patterned via photolithography to open a window on the SiN waveguides. Alignment markers, prepared by metal lift-off after photolithography and electron beam evaporation, were used for accurate placement of the opened windows on the SiN waveguides. Next, GO films were coated on the chip using the coating method mentioned above and patterned via a lift-off process. As compared with the drop-casting method that produces a GO film thickness of about 0.5 μm and a minimum size of about 1.3 mm for each step [49], the combination of our GO coating method with photolithography and lift-off allows precise control of the film placement (deviation < 20 nm), size (down to 100 nm) and thickness (with an ultrahigh resolution of ≈2 nm). The precise deposition and patterning control, along with the large area coating capability, is critical for large-scale, highly precise and cost-effective integration of 2D layered GO films on-chip. Apart from allowing precise control of the size and placement of the GO films that are in contact with the SiN waveguides, the patterned GO films also enabled us to test the performance of devices having a shorter length of GO film but with higher film thicknesses, which provides more flexibility to optimize the device performance with respect to FWM CE and bandwidth.

*2.2 Device characterization*

**Figure 2a** shows a microscope image of a SiN waveguide patterned with 10 layers of GO, which illustrates the high transmittance and good morphology of the GO films. **Figure 2b** presents a scanning electron microscopy (SEM) image of a GO film with up to 5 layers of GO monolayers, clearly showing the layered film structure. **Figure 2c** shows the measured Raman spectra of a SiN chip without GO and with 10 layers of uniformly coated GO films. The successful integration of GO films is confirmed by the presence of the representative D (1345 cm$^{-1}$) and G (1590 cm$^{-1}$) peaks of GO [32, 41]. **Figure 2d** plots the GO film thickness as a function of GO layer number measured by atomic force microscopy. The plots show the average of measurements on three samples and the error bars reflect the variations. The GO film thickness shows a nearly linear relationship with the layer number, with a thickness of ≈2 nm on average for each layer.

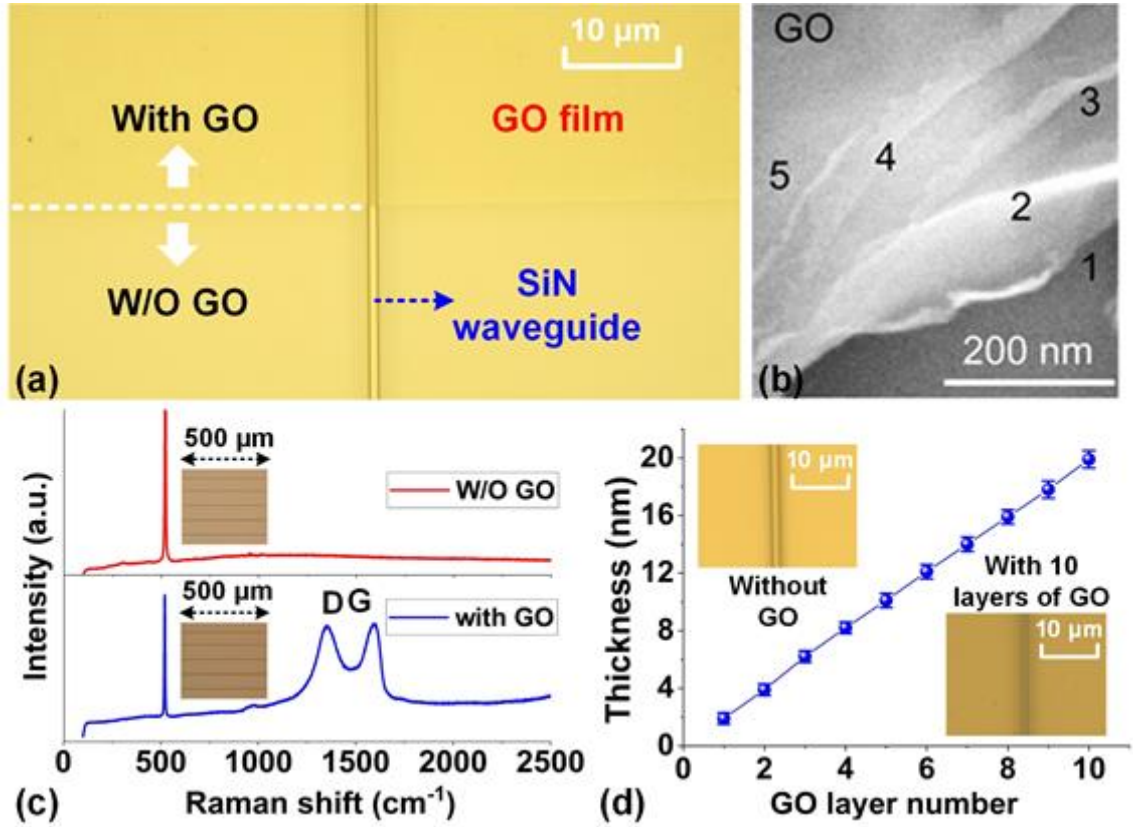

**Figure 2.** (a) Microscope image of a SiN waveguide patterned with 10 layers of GO. (b) SEM image of a GO film including 5 layers of GO. The numbers refer to the number of layers for that part of the image. (c) Raman spectra of a SiN chip without GO and with 10 layers of GO. Insets show the corresponding microscope images. (d) Measured GO film thickness versus layer number. Insets show the microscope images of an uncoated SiN waveguide and the same waveguide coated with 10 layers of GO.

We fabricated and tested two types of GO-coated SiN waveguides: the first with either 1 or 2 layers of uniformly coated GO films and the second with 5 or 10 layers of patterned GO films. The length of the SiN waveguides was 20 mm, which was the same as the GO coating length for the uniformly coated devices. For the patterned devices, the GO films were coated at the beginning of the SiN waveguides and the coating length was 1.5 mm. **Figure 3a** depicts the insertion loss of the GO-coated SiN waveguides measured using a transverse electric (TE) polarized continuous-wave (CW) light with a power of 5 dBm. We employed lensed fibers to butt couple the CW light into and out of the SiN waveguides with inverse-taper couplers at both ends. The butt coupling loss was ≈5 dB per facet, corresponding to 0-dBm CW power coupled into the waveguides.

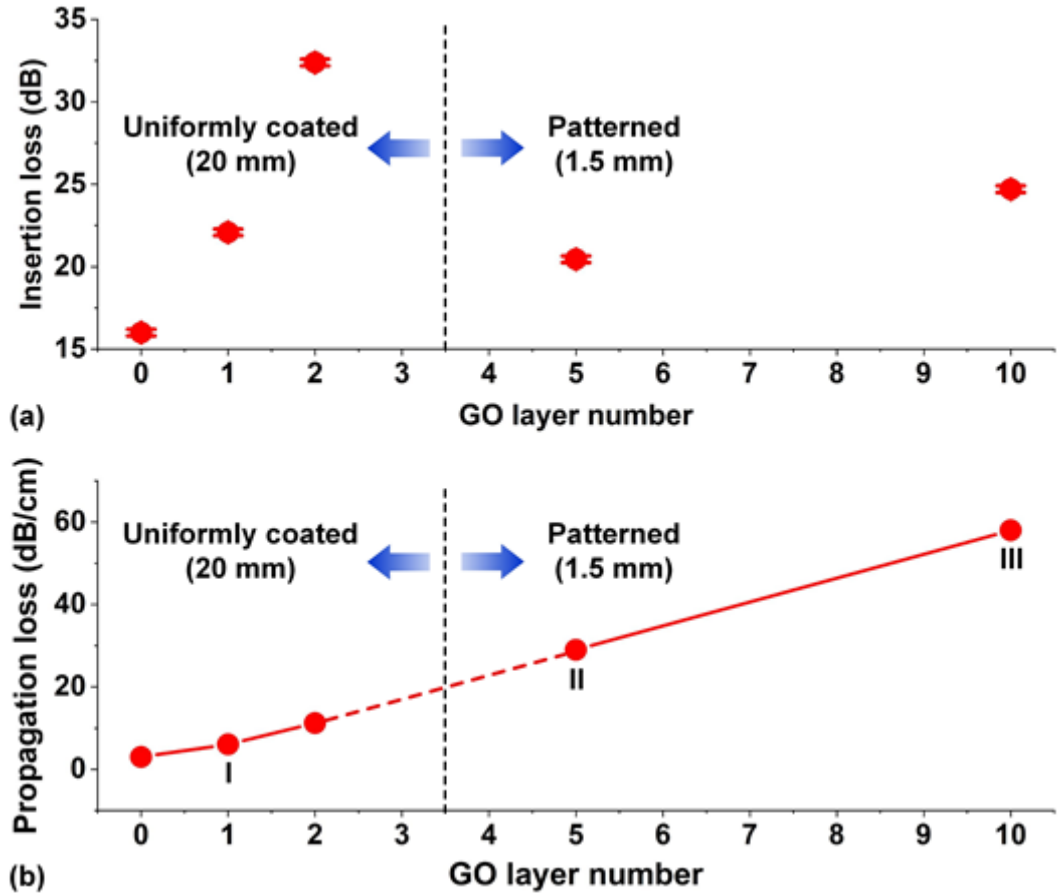

**Figure 3.** (a) Measured insertion loss of SiN waveguides with uniformly coated and patterned GO films. (b) Propagation loss of the hybrid waveguides extracted from (a). The slope rates (SRs) of the curve at points I, II and III are 3.1, 4.7 and 5.5 dB/cm/layer, respectively. In (a) and (b), the results for the bare SiN waveguide (i.e., the GO layer number is 0) are also shown for comparison.

**Figure 3b** shows the propagation loss of the SiN waveguides coated with different numbers of GO layers extracted from **Figure 3a**. The propagation loss of the bare SiN waveguides was ≈3.0 dB/cm, which was obtained from cutback measurements of SiN waveguides with the same geometry but different lengths. The propagation loss of the SiN waveguides with a monolayer of GO was ≈6.1 dB/cm, corresponding to an excess propagation loss of ≈3.1 dB/cm induced by the GO film. This is about a factor of 3 higher than reported for doped silica waveguides and mainly results from the higher mode overlap in the SiN waveguide reported here versus the much larger buried waveguides in doped silica [32, 50]. The loss reported here is also about 2 orders of magnitude smaller than SiN waveguides coated with graphene [31], reflecting the low material absorption of GO and its strong potential for the implementation of high-performance nonlinear photonic devices. In contrast to graphene that has a metallic behavior (e.g., high electrical and thermal conductivity) with zero bandgap, GO is a dielectric that has a large bandgap of 2.1−2.4 eV [41, 48], which results in low linear light absorption in spectral regions below the bandgap. In theory, GO films with a bandgap > 2 eV should have negligible absorption at near-infrared wavelengths. We therefore infer that the linear loss of the GO films is mainly due to light absorption from localized defects as well as scattering loss stemming from film unevenness and imperfect contact between the different layers. We note that the linear loss of the GO films is not a fundamental property. Therefore, by optimizing our GO synthesis and coating processes, such as using GO solutions with reduced flake sizes and increased purity, it is anticipated that the loss of our GO films can be further reduced. In **Figure 3b**, we label the slope rates of the curve at 1, 5 and 10 layers of GO, where we see that the propagation loss of the hybrid waveguides increases with GO layer number super linearly. This is a result of an increase in the contributions just outlined, as reported previously [32, 50].

**3. FWM experiment**

**Figure 4** shows the experimental setup used to measure FWM in the GO-coated SiN waveguides. Two CW tunable lasers separately amplified by erbium-doped fiber amplifiers (EDFAs) were used as the pump and signal sources, respectively. In each path, there was a polarization controller (PC) to ensure that the input light was TE-polarized. The pump and signal were combined with a 3-dB fiber coupler before being coupled into the hybrid waveguide as device under test (DUT). A charged-coupled device (CCD) camera was set above the DUT for coupling alignment. An optical isolator was employed to prevent the reflected light from damaging the laser source. The signal output from waveguide was sent to an optical spectrum analyzer (OSA) with a variable optical attenuator (VOA) to prevent high-power damage.

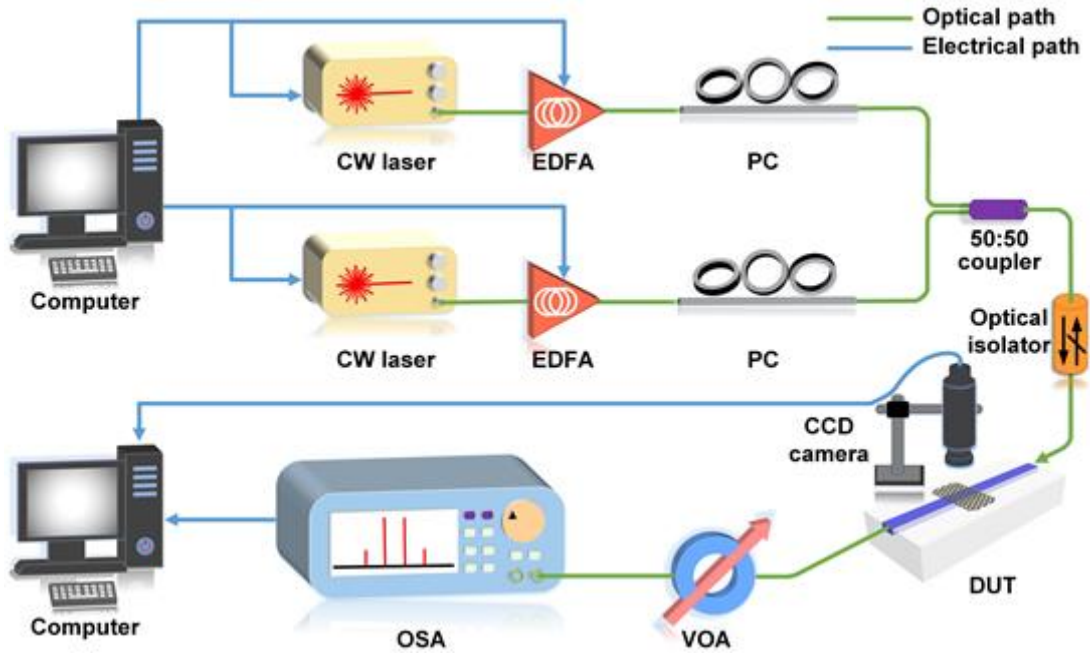

Fig. 4. Experimental setup for testing FWM in the GO-coated SiN waveguides. EDFA: erbium-doped fibre amplifier, PC: polarization controller, DUT: device under test, OSA: optical spectrum analyser, CCD: charged-coupled device and VOA: variable optical attenuator.

**Figure 5a-i** shows the experimental FWM optical spectra for the SiN waveguides uniformly coated with 1 and 2 layers of GO, together with the FWM spectrum of the bare SiN waveguide. For comparison, we kept the same power of 23 dBm for both the pump and signal before the input of the waveguides, which corresponded to 18 dBm power for each coupled into the waveguides. The difference among the baselines of the spectra reflects the difference in waveguide propagation loss for different samples. It can be seen that although the hybrid waveguide with 1 layer of GO film had an additional propagation loss of ≈7.1 dB, it clearly shows enhanced idler output powers as compared with the bare SiN waveguide. The CE (defined as the ratio of the output power of the idler to the input power of the signal, i.e., $P_{out, idler}$ / $P_{in, signal}$) of the SiN waveguides without GO and with 1 layer of GO were ≈-65.7 dB and ≈-58.4 dB, respectively, corresponding to a CE enhancement of ≈7.3 dB for the hybrid waveguide. In contrast to the positive CE enhancement for the hybrid waveguide with 1 layer of GO, the change in CE for the hybrid waveguide with 2 layers of GO was negative. This mainly resulted from the increase in propagation loss with GO layer numbers, as noted in **Figure 3b**.

**Figure 5a-ii** shows the FWM spectra of the SiN waveguides with 5 and 10 layers of patterned GO films. The coupled CW pump and signal power (18 dBm for each) was the same as that in **Figure 5a-i**. The SiN waveguides with patterned GO films also had an additional insertion loss as compared with the bare waveguide, while the results for both 5 and 10 GO layers show enhanced idler output powers. In particular, there is a maximum CE enhancement of ≈9.1 dB for the SiN waveguide patterned with 5 layers of GO, which is even higher than that for the uniformly coated waveguide with 1 layer of GO. This reflects the trade-off between FWM enhancement (which dominates for the patterned devices with a short GO coating length) and loss (which dominates for the uniformly coated waveguides with a much longer GO coating length) in the GO-coated SiN waveguides (see Section 4).

**Figure 5b** shows the measured CE versus pump power for the uniformly coated and patterned devices, respectively. The plots show the average of three measurements on the same samples and the error bars reflect the variations, showing that the measured CE is repeatable. As the pump power was increased, the measured CE increased linearly with no obvious saturation for the bare SiN waveguide and all the hybrid waveguides, indicating the low TPA of both the SiN waveguides and the GO films. For the bare waveguide, the dependence of CE versus pump power shows a nearly linear relationship, with a slope rate of about 2 for the curve as expected from classical FWM theory [17]. For the GO-coated waveguides, the measured CE curves have shown slight deviations from the linear relationship with a slope rate of 2, particularly at high light powers. This is a reflection of the change in GO material properties with light power (see Section 4).

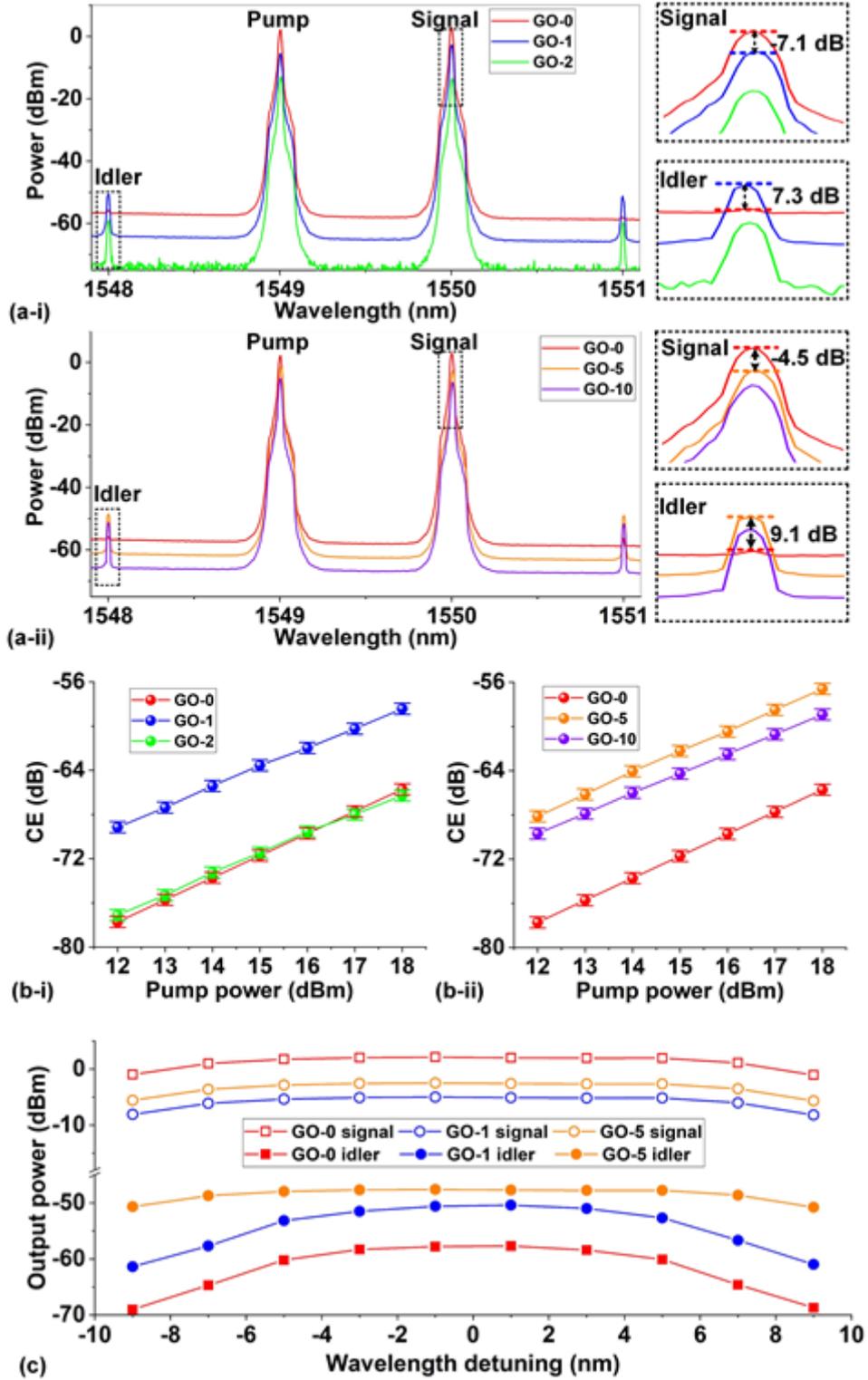

**Figure 5**. FWM experimental results. (a) FWM optical spectra. Insets show zoom-in view around the signal and idler. (b) CE versus pump power coupled into the waveguides. In (a) and (b), (i) shows the results for SiN waveguides uniformly coated with 1 and 2 layers of GO and (ii) shows the results for SiN waveguides patterned with 5 and 10 layers of GO. (c) Power variations of the output signal and idler when the pump wavelength was fixed at 1550 nm and the signal wavelength was detuned around 1550 nm for the uniformly coated device with 1 layer of GO and the patterned device with 5 layers of GO. In (a)−(c), the corresponding results for the bare SiN waveguide (GO-0) are also shown for comparison. In (b) and (c), the solid lines are merely guides to the eye.

**Figure 5c** shows the output signal/idler power versus wavelength detuning (i.e., wavelength spacing between pump and signal) for the uniformly coated device with 1 layer of GO and the patterned device with 5 layers of GO. The results for the bare SiN waveguide are also shown for comparison. The coupled pump power was 18 dBm, with the pump wavelength fixed at 1550 nm and the signal wavelength detuned from 1540 nm to 1560 nm. For all devices the output signal powers decreased with increasing wavelength detuning, which was caused by gain roll-off of the EDFA. The output idler power, on the other hand, decreased even more rapidly with wavelength detuning, and this was predominantly a result of decreased phase-matching. As compared with the bare and uniformly coated SiN waveguides, the patterned devices showed a much broader FWM bandwidth with higher idler power on both edges, reflecting a wider FWM phase matching bandwidth for a shorter length of GO films as expected.

## 4. Theoretical analysis and discussion

### 4.1 FWM theory

We used the theory from Refs. [32, 56, 57] to model the FWM process in the GO-coated SiN waveguides. Assuming negligible depletion of the pump and signal powers due to the generation of the idler, the coupled differential equations for the degenerate FWM process can be expressed as [58, 59]

$$\frac{dA_p(z)}{dz} = -\frac{\alpha_p}{2} A_p(z) + j\gamma_p \left[ \left| A_p(z) \right|^2 + 2|A_s(z)|^2 + 2|A_i(z)|^2 \right] A_p(z)$$

$$+ j2\gamma_p A_p^*(z) A_s(z) A_i(z) exp(j\Delta\beta z) \tag{1}$$

$$\frac{dA_s(z)}{dz} = -\frac{\alpha_s}{2} A_s(z) + j\gamma_s \left[ |A_s(z)|^2 + 2\left| A_p(z) \right|^2 + 2|A_i(z)|^2 \right] A_s(z)$$

$$+ j\gamma_s A_i^*(z) A_p^2(z) exp(-j\Delta\beta z) \tag{2}$$

$$\frac{dA_i(z)}{dz} = -\frac{\alpha_i}{2} A_i(z) + j\gamma_i \left[ |A_i(z)|^2 + 2\left| A_p(z) \right|^2 + 2|A_s(z)|^2 \right] A_i(z)$$

$$+ j\gamma_i A_s^*(z)A_p^2(z)exp(-j\Delta\beta z) \qquad (3)$$

where $A_{p,s,i}$ are the amplitudes of the pump, signal and idler waves along the z axis, which we define as the light propagation direction, $\alpha_{p,s,i}$ are the linear losses, $\Delta\beta = \beta_s + \beta_i - 2\beta_p$ is the linear phase mismatch, with $\beta_{p,s,i}$ denoting the propagation constants of the pump, signal and idler waves, and $\gamma_{p,s,i}$ are the waveguide nonlinear parameters. In our case, where the wavelength detuning range was small ($\leq 10$ nm), we assumed that the linear loss and the nonlinear parameter are constant, i.e., $\alpha_p = \alpha_s = \alpha_i = \alpha$, $\gamma_p = \gamma_s = \gamma_i = \gamma$.

Since the bandgaps of SiN (5 eV [2]) and GO (2.1−2.4 eV [41, 48]) are much larger than the TPA bandgap (1.6 eV) in the telecommunications band, we neglected nonlinear loss induced by TPA of SiN and GO in **Eqs. (1)** − **(3)**. In our previous Z-scan measurements [42, 46], we observed saturable absorption (SA) behavior (with loss decreasing with light power – a trend that is opposite to TPA) for the GO films as a result of using optical pulses with higher peak powers (> 10 W). In our FWM experiment, we did not observe any SA phenomenon for the hybrid waveguides. This is probably because the peak powers of the CW light were much lower (< 0.15 W, the power in the GO films was even lower given the mode overlap with GO).

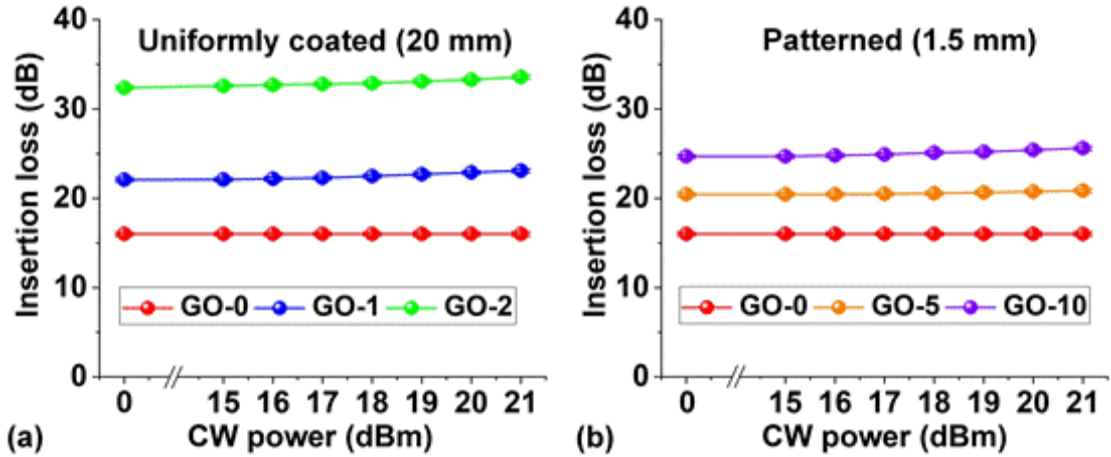

**Figure 6.** Measured insertion loss of SiN waveguides with (a) uniformly coated and (b) patterned GO films versus input CW power. The results for the bare SiN waveguide (GO-0) are also shown for comparison.

**Figures 6a**, **b** depict the insertion loss of the GO-coated SiN waveguides versus input CW power (after excluding the butt coupling loss). There was small but observable increase in the insertion loss with input CW power for the GO-coated waveguides. In contrast, we could not observe any obvious changes for the bare (uncoated) waveguide. This indicates that the change in the insertion loss of the hybrid waveguides was induced by the GO films. We also note that the power-induced loss changes were not permanent – when the CW power was reduced the measured insertion loss recovered to that at low power in **Figure 3a**, with the measured insertion loss being repeatable. This phenomenon is similar to that observed from GO-coated doped silica waveguides and can be attributed to the photo-thermal changes of GO films [50, 60]. The absorbed CW power generated heat and increased the temperature of the hybrid waveguides, which temporarily modified some OFGs in the GO films. The photo-thermal induced changes in the OFGs could modify both the linear loss and $n_2$, and depend on the average CW power. This is distinct from TPA-induced loss that occurs instantaneously and depends on peak power. Since the time response for photo-thermal changes is slow, we accounted for the power-dependent loss of the GO films by using the measured loss versus CW power in **Figures 6a, b** to calculate $\alpha$ of the hybrid waveguides in **Eqs. (1) − (3)**. Note that there were the same overall CW powers coupled into the waveguides (assuming the idler power could be neglected) for a single CW light with 15−21 dBm power in **Figures 6a, b** and two CW lights (pump and signal with the same power) with 12−18 dBm power for each in **Figures 5a, b**.

The dispersions $\beta_{p,s,i}$ in **Eqs. (1) − (3)** were calculated by Lumerical FDTD commercial mode solving software using the refractive index $n$ and extinction coefficient $k$ of layered GO films measured by spectral ellipsometry. By numerically solving **Eqs. (1) – (3)**, the FWM CE was calculated via

$$\text{CE (dB)} = 10 \times \log_{10}[|A_i(L)|^2/|A_s(0)|^2] \tag{4}$$

where $L$ is the length of the SiN waveguide (i.e., 20 mm). For the patterned devices, the waveguides were divided into bare SiN (without GO films) and hybrid (with GO films) segments with different $\alpha$, $\gamma$ and $\beta_{p,s,i}$. The FWM differential equations in **Eqs. (1) − (3)** were solved for each segment, with the output from the previous segment as the input for the subsequent segment.

*4.2 FWM analysis*

**Figure 7a** shows the experimental and theoretically calculated CE as a function of wavelength detuning for the bare SiN waveguide, the uniformly coated device with 1 layer of GO and the patterned device with 5 layers of GO. The measured CE values, obtained from the raw experimental results in **Figure 5c** after accounting for the EDFA gain roll-off, show good agreement with theory from **Eqs. (1) − (4)**. The patterned device has not only a higher CE, but also a broader FWM bandwidth. According to Ref. [16], the FWM bandwidth can be approximated by

$$\tilde{} \text{BW} \approx \left[\frac{4\pi}{\beta_2 L}\right]^{\frac{1}{2}} \tag{5}$$

where $\beta_2$ is group-velocity dispersion (GVD) and $L$ is the interaction length. In **Eq. (5)**, the FWM bandwidth is inversely proportional to the square root of the product of $\beta_2$ and $L$, i.e., reducing $\beta_2$ or $L$ increases the FWM bandwidth. The increased bandwidth for the patterned device resulted mainly from the shorter GO coating length. A small contribution arose from better phase matching due to a slightly enhanced anomalous dispersion for the hybrid waveguides (with $\beta_2 \approx -1.05 \times 10^{-25}$ s$^2$ m$^{-1}$ for the hybrid waveguide with 10 layers of GO calculated by FDTD simulations) versus the uncoated SiN waveguide (with $\beta_2 \approx -1.0 \times 10^{-25}$ s$^2$ m$^{-1}$). These effects complement the strong Kerr nonlinearity of the layered GO films, which is the dominant cause of enhanced FWM in the hybrid waveguides.

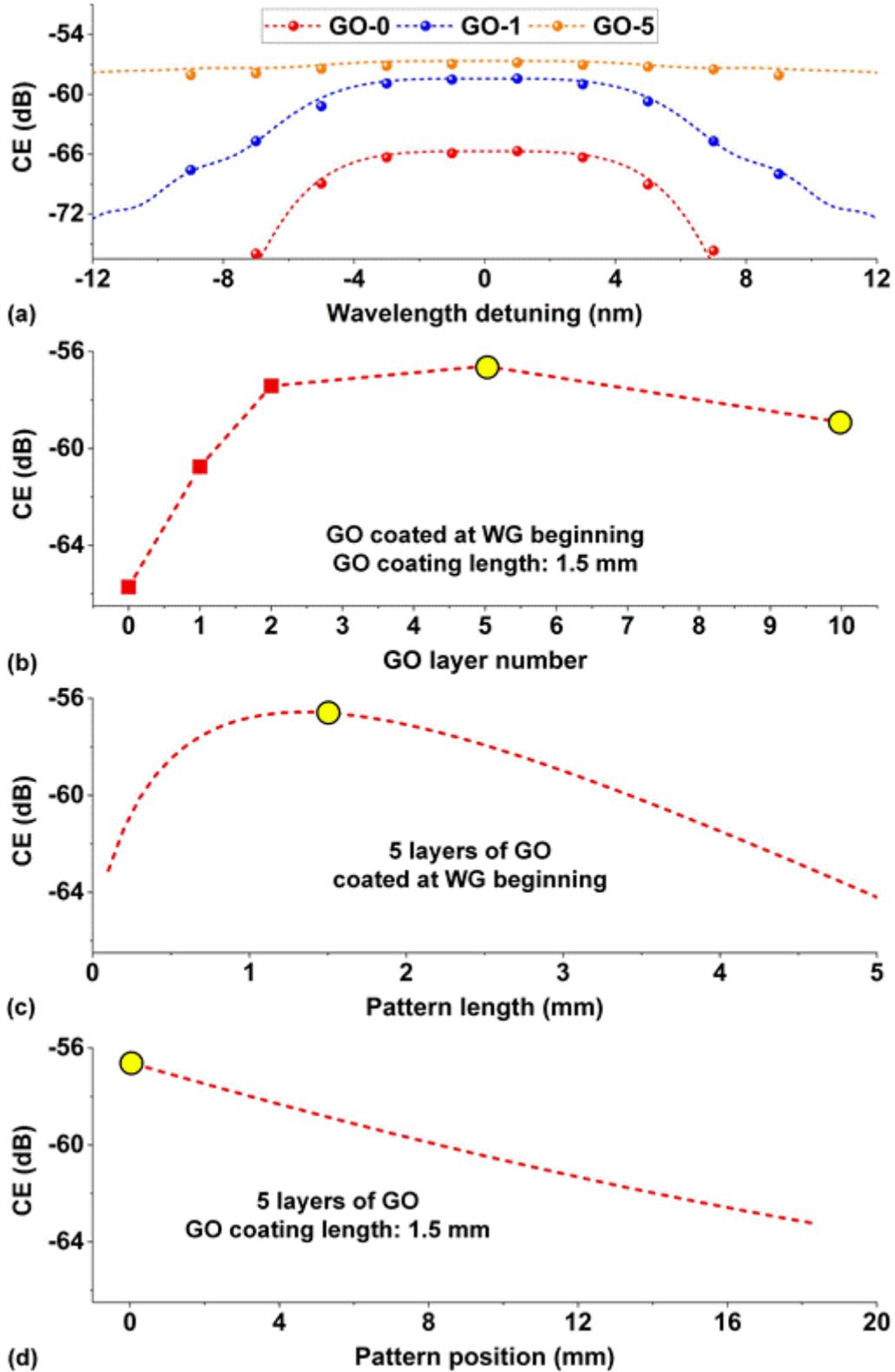

**Figure 7**. (a) Measured (data points) and fit (dashed curves) CE versus wavelength detuning for the bare SiN waveguide (GO-0), the uniformly coated device with 1 layer of GO (GO-1) and the patterned device with 5 layers of GO (GO-5). (b)−(d) Calculated CE as functions of GO layer number, coating length and coating position, respectively. WG: waveguide. The yellow data points show the experimental results. For comparison, the coating length in (b) and (d) is 1.5 mm, the GO layer number in (c) and (d) is 5, and the GO coating position in (b) and (c) is at waveguide beginning.

**Figures 7b−d** show the CE of the hybrid waveguides calculated from **Eqs. (1) − (4).** We compared the CE performance of the hybrid waveguides while varying three parameters of the GO films including the layer number, coating length and coating position. In each subfigure, we only changed one parameter, keeping the other two constant. **Figure 7b** compares the CE of the hybrid waveguides with four different numbers of GO layers (i.e., 1, 2, 5, 10), where we see that the hybrid waveguide with an intermediate number of GO layers has the maximum CE. This reflects the trade-off between $\gamma$ and loss in the hybrid waveguides, which both increase with GO layer number. **Figure 7c** plots the CE of the hybrid waveguides as a function of GO coating length. Similar to the trend with GO layer number, the maximum CE is obtained for intermediate GO coating lengths, reflecting a trade-off where the Kerr nonlinearity enhancement dominates for short GO coating lengths while the loss increase dominates for longer lengths. **Figure 7d** shows the CE of the hybrid waveguides as a function of GO coating position. In contrast to the trend in **Figures 7b, c**, the hybrid waveguide with GO films at the beginning (i.e., pattern position = 0) has the greatest CE. This is expected since the pump power in the GO film is highest at the start of the waveguide and decreases as the GO segment moves further along the waveguide. Clearly this effect gets smaller with decreasing propagation loss of the bare waveguide, being much lower (< 0.5 dB) for doped silica waveguides (with a propagation loss of 0.24 dB/cm [32] ) versus the SiN waveguides (with a propagation loss of ≈3 dB/cm) studied here.

*4.3 Nonlinear parameter ( $\gamma$ ) of the hybrid waveguides and $n_2$ of the GO films*

**Figure 8** shows the nonlinear parameter $\gamma$ of the hyrbid waveguides with different numbers of GO layers obtained from fitting theory to experiment, both at high (18 dBm) and low (12 dBm) pump powers. As expected, $\gamma$ increases with the GO layer number. In particular, for the SiN waveguides with 10 layers of GO, $\gamma$ is about two orders of magnitude higher than the bare SiN waveguide. The very small change in $\gamma$ with power mainly arises from a corresponding change in $n_2$ of the GO films.

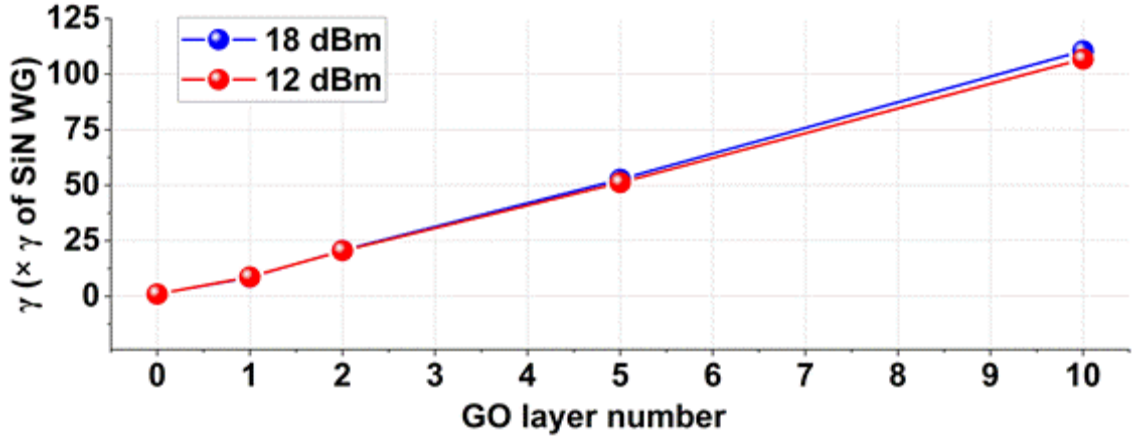

**Figure 8.** Nonlinear parameter (γ) of GO-SiN hybrid waveguides (normalized to γ of the bare SiN waveguide) versus number of GO layers for high (18 dBm) and low (12 dBm) pump powers. WG: waveguide.

Based on the values for γ of the hybrid waveguides obtained from the FWM experiments, we calculated the Kerr coefficient ($n_2$) of the layered GO films using [32, 47]:

$$\gamma = \frac{2\pi}{\lambda} \frac{\iint_D n_0^2(x,y) n_2(x,y) S_z^2 dxdy}{\left[\iint_D n_0(x,y) S_z dxdy\right]^2} \tag{6}$$

where λ is the pump wavelength, $D$ is the integral of the optical fields over the material regions, $S_z$ is the time-averaged Poynting vector calculated using COMSOL Multiphysics, $n_0(x, y)$ and $n_2(x, y)$ are the linear refractive index and $n_2$ profiles over the waveguide cross section, respectively. This work was performed in the regime close to degeneracy where the three FWM frequencies (pump, signal, idler) were close together compared with any dispersion in $n_2$ [32]. We therefore used $n_2$ instead of the more general third-order nonlinearity ($\chi^{(3)}$) in our analysis. The values of $n_2$ for silica and silicon nitride used in our calculations were $2.60 \times 10^{-20}$ m²/W [2] and $2.61 \times 10^{-19}$ m²/W, respectively, the latter obtained by fitting the experimental results for the bare SiN waveguide. Note that γ in **Eq. (6)** is an effective nonlinear parameter weighted not only by $n_2(x, y)$ but also by $n_0(x, y)$ in the different material regions, which is more accurate for high-index-contrast hybrid waveguides studied here as compared with the theory in Refs. [32, 61].

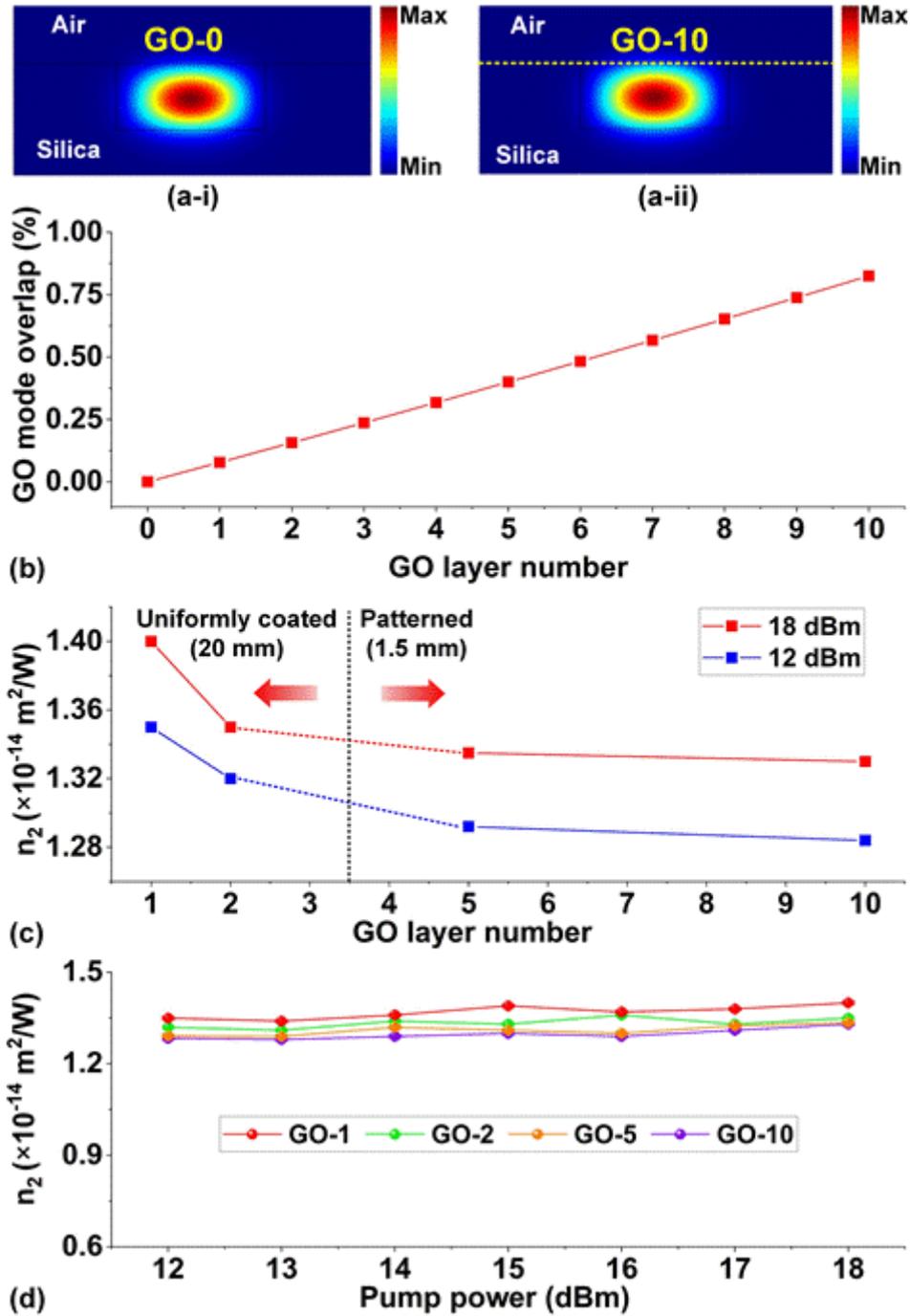

**Figure 9.** (a) TE mode profiles for the SiN waveguides (i) without GO and (ii) with 10 layers of GO. (b) Mode overlap with GO versus GO layer number for the hybrid waveguides. (c) $n_2$ of layered GO films versus GO layer number at fixed pump powers of 12 dBm and 18 dBm. (d) $n_2$ of layered GO films versus pump power for the hybrid waveguides with different numbers of GO layers.

**Figure 9a** shows the TE mode profiles for the SiN waveguides without GO and with 10 layers of GO. The mode overlap with GO films versus GO layer number is shown in **Figure 9b**, which was calculated by integrating the time-averaged Poynting vectors for different material regions. Most of the power is confined to the SiN waveguide (88.3% and is constant within 0.2%) and the mode overlap with the GO films is small (< 1%). This is not surprising given the difference in volume between the bulk SiN waveguide and the ultrathin 2D GO film. The mode overlap with GO film increases with GO layer number, leading to an increased loss and $\gamma$ for the hybrid waveguide with thicker GO films.

**Figure 9c** shows $n_2$ versus layer number for the GO films at fixed pump powers of 12 dBm and 18 dBm. The $n_2$ values, although slightly lower than graphene [62, 63], are nonetheless over four orders of magnitude higher than SiN and agree reasonably well with our previous measurements [32, 42, 46, 47]. Such a high $n_2$ for the GO films highlights their strong Kerr nonlinearity not only for FWM but also other third-order ($\chi^{(3)}$) nonlinear processes such as SPM and cross phase modulation (XPM), and possibly even enhancing $\chi^{(3)}$ for THG and stimulated Raman scattering [13, 24, 46, 64]. We observe that $n_2$ (both at 12 dBm and 18 dBm) decreases with GO layer number, similar to the trend observed for layered $WS_2$ films measured by a spatial-light system [65]. In our case, this was probably a result of an increase in inhomogeneous defects within the GO layers as well as imperfect contact between the multiple GO layers. We also note that the rate of decrease in $n_2$ with GO layer number decreases for thicker GO films, reflecting the transition of the GO film properties towards bulk properties, with a thickness independent $n_2$.

In **Figure 9d**, we plot $n_2$ for the GO films as a function of pump power coupled into the hybrid waveguides, which shows a very slight change in $n_2$ with power that is reversible. Unlike the monotonic decrease in $n_2$ with GO layer number that we observe, the power dependent change in $n_2$ shows very slight oscillations. This is similar to that observed from FWM in GO-coated MRRs [47], and can be attributed to the power-sensitive (reversible) photo-thermal changes of GO [50, 66] as well as self-heating and thermal dissipation in the multiple GO layers. The power-dependent change in $n_2$ we obtained here is much smaller than that from GO-coated MRRs [47], which is perhaps not surprising since the light intensity in MRRs is much higher due to the resonant enhancement of the optical field.

We verified that all measurements (insertion loss and CE) were repeatable, reflecting the fact that no permanent changes in the material properties of the GO films occured. Previously [42, 43, 67, 68], we demonstrated that the material properties of GO can be permanently modified by direct laser writing with high power femtosecond laser pulses. This is distinct from the non-permanent photo-thermal changes we observe here.

**Table I. Performance comparison of doped silica and SiN waveguides integrated with 2D layered GO films.**

| | $n_0$ [a] of WG | Waveguide dimension ($\mu$m) | $PL_{WG}$ [b] (dB/cm) | $EPL_{GO-1}$ [c] (dB/cm) | $\gamma_{WG}$ [d] ($W^{-1}m^{-1}$) | $\gamma_{hybrid}$ [e] ($W^{-1}m^{-1}$) | $n_2$ of GO ($m^2$/W) | Ref. |
|---|---|---|---|---|---|---|---|---|
| Doped silica | 1.66 | Width: 2.00 Height: 1.50 | 0.24 | 1.0 | 0.28 | GO-1: 0.61 GO-10: n/a [f] | $1.5 \times 10^{-14}$ | [32] |
| SiN | 1.99 | Width: 1.60 Height: 0.66 | 3.0 | 3.1 | 1.51 | GO-1: 13.14 GO-10:167.14 | $1.28 \times 10^{-14}$ $\sim 1.41 \times 10^{-14}$ | This work |

a) $n_0$ of WG: linear refractive indices of the bare waveguides.
b) $PL_{WG}$: propagation loss of the bare waveguides without GO films.
c) $EPL_{GO-1}$: excess propagation loss induced by GO for the hybrid waveguides with 1 layer of GO.
d) $\gamma_{WG}$: nonlinear parameters of the bare waveguides.
e) $\gamma_{hybrid}$: nonlinear parameters of the hybrid waveguides with 1 layer (GO-1) and 10 (GO-10) layers of GO.
f) Only hybrid waveguides with 1-5 layers of GO were characterized.

Finally, we compare these results with a previous demonstration of enhanced FWM in doped silica waveguides integrated with layered GO films [32]. **Table I** compares relevant parameters for doped silica and SiN waveguides incorporated with 2D GO films, where we see that the two waveguides were quite different. For this work, the excess propagation loss in the hybrid SiN waveguides induced by the GO film was much higher due to the significantly increased mode overlap with the GO film. On the other hand, this also resulted in a significantly increased $\gamma$ for the GO-SiN hybrid waveguides. Mode overlap is an important factor for optimizing the trade-off in nonlinear optical performance between the Kerr nonlinearity and loss when integrating 2D layered GO films onto integrated photonic devices. According to our simulations, the FWM CE can be further improved by redesigning the cross section of the SiN waveguide to optimize the mode overlap, particularly for SiN waveguides having a lower height (i.e., SiN film thickness) of < 400 nm. This is significant, given the stress-induced cracking observed for thick SiN films [69]. In contrast to the doped silica waveguides that employed only uniformly coated

GO films, here we find that the use of patterned GO films can result in a more significant improvement in FWM CE due to a better balance between loss and Kerr nonlinearity as well as a much broader FWM bandwidth. Finally, there is significant potential to reduce the intrinsic linear loss of the GO films, which is not fundamental as it is for graphene, and this represents the greatest opportunity to improve the nonlinear device performance.

**5. Conclusion**

We demonstrate improved FWM efficiency in SiN waveguides integrated with 2D layered GO films arising from an enhanced Kerr nonlinearity. SiN waveguides with both uniformly coated and patterned GO films are fabricated with precise control of the film thickness, placement and coating length. We perform FWM measurements for samples with different numbers of GO layers and at different pump powers, achieving up to ≈7.3 dB CE enhancement for a uniformly coated device with 1 layer of GO and ≈9.1 dB for a patterned device with 5 layers of GO. As compared with the uniformly coated devices, both improved FWM CE and bandwidth are achieved for the patterned devices. The influence of pattern length and position on the FWM performance are also analysed. By fitting the experimental results with theory, we obtain up to 100 times improvement in the nonlinear parameter for the hybrid waveguides as well as the change of GO's third-order nonlinearity with layer number and pump power. This work demonstrates that integrating 2D layered GO films onto SiN devices can effectively transform SiN into a viable and highly performing nonlinear photonic platform, which we believe could play an important role in integrated nonlinear optics well beyond the FWM process studied here.

**Acknowledgements**

Y. Qu and J. Wu contribute equally to this work. This work was supported by the Australian Research Council Discovery Projects Programs (No. DP150102972 and DP190103186), the Swinburne ECR-SUPRA program, the Industrial Transformation Training Centers scheme (Grant No. IC180100005), and the Beijing Natural Science Foundation (No. Z180007). The authors also acknowledge the Swinburne Nano Lab for the support in device fabrication and characterization.